\documentstyle[11pt,a4]{article}

\begin{document}

\begin{titlepage}
\begin{flushright}
IJS-TP-98/08\\
TECHNION-PH-98-10\\
hep-ph/9805461\\
17 Jul 98\\ 
\end{flushright}

\vspace{.5cm}

\begin{center}
{\Large \bf Resonant and nonresonant contributions to the weak $D\to Vl^+l^-$ decays\\}

\vspace{.5cm}

{\large \bf S. Fajfer$^{a}$, S. Prelov\v sek$^{a}$ and
P. Singer$^{b}$\\}

\vspace{1cm}
{\it a) J. Stefan Institute, Jamova 39, P. O. Box 300, 1001 Ljubljana, 
Slovenia}\vspace{.5cm}

{\it b) Department of Physics, Technion - Israel Institute  of Technology, 
Haifa 32000, Israel\\}
\end{center}

\vspace{1cm}

\centerline{\large \bf ABSTRACT}

\vspace{0.5cm}  
The Cabibbo suppressed decays $D\to Vl^+l^-$ ($V$ is light vector meson) present in principle  the 
opportunity to observe the 
short distance FCNC transition $c\to ul^+l^-$, which is sensitive 
to physics 
beyond the Standard Model. 
We analyze these as well as the Cabibbo allowed  $D \to V l^+ l^- $ 
decays within the Standard Model, where in addition to the short distance 
dynamics also the long distance dynamics is present. 
    The long distance contribution is induced by the effective nonleptonic weak Lagrangian accompanied by the emission of a virtual photon, which occurs resonantly via conversion from a vector meson $\rho^0$, $\omega $ or $\phi$ or nonresonantly as direct emission from a $D$ meson. 
We  calculate the branching ratios for all 
$D\to Vl^+l^- $ decays 
using the model, which combines heavy quark symmetry and chiral 
perturbation theory. The short distance contribution due to $c\to ul^+l^-$ transition, which is present only in the Cabibbo suppressed decays,  is found to be three orders of magnitude smaller than  the long distance contribution. The branching ratios well above $10^{-7}$ for  Cabibbo suppressed decays could signal new physics. The most frequent 
decays   are the Cabibbo allowed decays, which are expected at the rates, that are not much lower than the present experimental upper limit: $D_s^+\to 
\rho^+\mu^+\mu^-$ is expected at the branching ratio of approximately $3\cdot 10^{-5}$, while $D^0\to\bar K^{*0}\mu^+\mu^-$ is expected at $1.7\cdot 10^{-6}$. \\

\end{titlepage}

\setlength {\baselineskip}{0.55truecm}

\setcounter{footnote}{1} 

\setcounter{footnote}{0}

{\bf I. INTRODUCTION}\\ 

In the charm sector phenomena like $D^0-\bar D^0$ mixing, 
CP-violation and  rare decay probabilities are small, which makes them good 
candidates as probes for new physics with small background from the
Standard Model \cite{bigi1,hew,pak}. In particular, 
decays of type $D\to Xl^+l^-$ were singled out \cite{SCHWARTZ,castro,BABU} 
as a possible 
good window to non-standard contributions of the flavour-changing neutral 
transition (FCNC) $c\to ul^+l^-$, at the $10^{-7}$ level for the branching ratios. 
This suggestion was 
prompted by the smallness of the short-distance (SD) $c\to 
ul^+l^-$ contribution within the Standard Model, which leads \cite{SCHWARTZ,BABU} to a branching ratio of only 
$10^{-9}$ for the inclusive process. Although QCD corrections to this 
process have not been calculated in detail yet, these are not expected to affect 
significantly the size of the $c\to ul^+l^-$ amplitude, as explained in the next 
section. Accordingly, one expects the 
hadronic exclusive decays induced by this SD transition to occur with branching 
ratios of the order of $10^{-10}$.

Further studies, which have considered the long-distance (LD) contribution 
to $D\to Pl^+l^-$ transitions ($P$ is light pseudoscalar) \cite{BABU,SZ} have concluded that these 
are larger than SD ones. The analysis of the LD contributions in 
$D^{+,0}\to\pi^{+,0}l^+l^-$ \cite{SZ} has shown these modes are expected to lead 
to branching ratios of the order of $10^{-6}$ in the resonance region and a few 
times $10^{-7}$ in the nonresonant region, thus practically invalidating their 
use for observing the $c\to ul^+l^-$ transition within the Standard model.

A similar situation holds in the case of $D\to V\gamma$ ($D\to P\gamma$ is forbidden), where long distance 
effects \cite{BGHP,FS,FPS} cause these modes to have branching ratios in the 
$10^{-7}-10^{-4}$ range. On the other hand, the SD component due to the magnetic 
electroweak penguin transition $c\to u\gamma$ is GIM suppressed and does not reach beyond the 
$10^{-9}$ range at most, despite of being considerably 
enhanced by QCD corrections \cite{BGHP,GHMW}. Thus, here again the LD effects mask the contribution 
of the SD $c\to u\gamma$ loop, except for very special circumstances, which were 
pointed out recently \cite{FS,BFO0}. 

We analyze LD and SD contributions to all $D \to V l^+ l^- $ 
decays within the Standard Model. The SD contribution due to $c\to ul^+l^-$ is present only in the 
Cabibbo suppressed decays $D^0 \to \rho^0 l^+ l^-$, $D^0 \to \omega l^+ l^-$ 
$D^0 \to \phi l^+ l^-$, $D^+ \to \rho^+ l^+ l^-$ and 
$D_s^+ \to K^{*+} l^+ l^-$. Our results should provide the appropriate theoretical 
background against which possible signals of new physics are searched for in these decays. Motivated by the experimental searches, we analyze also the  Cabibbo allowed decays 
($D^0\to K^{*0}l^+l^-$ and $D_s^+\to \rho^+l^+l^-$), which are the best candidates for their early detection, and the doubly Cabibbo 
suppressed decays ($D^+\to K^{*+}l^+l^-$ and $D^0\to K^{*0}l^+l^-$). Here  the 
signals from new physics are not expected from the theoretical models usually 
considered. 

On the experimental side, so far there are only upper bounds on the branching ratios of  $D\to 
Vl^+l^-$ decays from E653 and CLEO \cite{exp,PDG}, in the range 
$10^{-3}-10^{-4}$, but these are expected to improve in the future.

In Sec. II we present the details of our  approach and we define the 
approximations used. In Sec. III we give the results of our calculations and we 
summarize in Sec. IV.\\ 

{\bf II. MODEL DESCRIPTION}\\  

{\bf A. Long distance contributions}\\

In this subsection we present the general framework used for calculating the long distance amplitudes, while the details of the model employed are given in subsection II C. 
The long distance contribution in $D\to Vl^+l^-$ decays is due to the effective nonleptonic weak Lagrangian, which induces the weak transition between the initial and final hadronic state. The weak transition has to be accompanied by the emission of a virtual photon, which finally decays into a lepton antilepton pair.
The  effective nonleptonic weak Lagrangian responsible for charm meson decays is
\begin{eqnarray}
\label{weak}
{\cal L}_{LD} =  -{G_F \over \sqrt{2}} V_{uq_i}V_{cq_j}^*  
~[ a_1 ({\bar u} q_i)^{\mu}
({\bar q_j}c )_{\mu}  +   a_2 ({\bar u} c)_{\mu} ({\bar q}_j q_i)^{\mu}], 
\end{eqnarray}
where $({\bar \psi}_1\psi_2)^\mu\equiv
{\bar \psi}_1\gamma^\mu(1-\gamma^5)\psi_2$, $q_{i,j}$ represent the fields of 
$d$ or $s$ quarks, $V_{ij}$ are the CKM matrix elements and $G_F$ is the 
Fermi constant. 
In our calculation we use $a_1 = 1.26$ and $a_2 = -0.55$ as 
found in \cite{bauertwo}, from an extensive application of (\ref{weak}) to the study of nonleptonic $D$ decays. 

The virtual photon emission from the hadronic states is taken in our approach to proceed through two different mechanisms:

\noindent
(i) In the {\it nonresonant mechanism}  the photon is emitted directly from the 
initial $D$ state.

\noindent
(ii) In the {\it resonant mechanism}, apart from the final vector meson $V$, an 
additional neutral vector meson  $V_0$ is produced, which converts to a photon 
through vector meson dominance (VMD). In this case, a nonleptonic weak decay 
$D\to VV_0$ is followed by the transition $V_0\to\gamma^*\to l^+l^-$, where 
$V_0$ is a short-lived vector meson $\rho^0$, $\omega$ or $\phi$.

 The evaluation 
of the matrix elements of the product of two currents (\ref{weak}) requires  nonperturbative techniques and we 
are forced to use some approximation. We have undertaken to use systematically the factorization approximation, where the matrix element of the product of two currents is approximated by 
\begin{eqnarray}
\label{factor}
\langle V \gamma | (\bar q_iq_j)^{\mu}(\bar q_kc)_{\mu} |D\rangle &=&
\langle  V | (\bar q_iq_j)^{\mu}|0\rangle\langle \gamma|(\bar q_kc)_{\mu} |D\rangle
\nonumber\\
&+&\langle  \gamma | (\bar q_iq_j)^{\mu}|0\rangle\langle V|(\bar q_kc)_{\mu} 
|D\rangle
\nonumber\\
&+&\langle  V \gamma | (\bar q_iq_j)^{\mu}|0\rangle\langle 0|(\bar q_kc)_{\mu} 
|D\rangle~.
 \end{eqnarray}
The first two terms are the spectator contributions, in the following 
denoted by $A_{Spec, \gamma}$ and $A_{Spec,V}$, respectively, and the third term 
is the weak annihilation contribution, denoted by $A_{Annih}$. Here  $\gamma$ denotes the virtual photon.  

\vspace{1.mm}
To calculate the matrix elements in (\ref{factor}) we use the  hybrid model, 
which combines 
heavy quark effective theory (HQET) and chiral Lagrangian \cite{BFO2}-\cite{casrep}, which has been successfully employed already for $D$ meson decays in several papers  \cite{BFO2}-\cite{cas}. 
A very  detailed description of the hybrid model and its previous applications is given in \cite{casrep}.   

 The relevant hadronic degrees of freedom for the present calculation are heavy pseudoscalar ($D$) 
and vector mesons ($D^*$) and light pseudoscalar ($P$) and vector ($V$) mesons. 
Within this approach the diagrams that contribute to the amplitudes 
$A_{Spec,\gamma}$, $A_{Spec,V}$  and $A_{Annih}$ (\ref{factor}) are shown in  
Figs. 1a, 1b 
and 1c, respectively. Different diagrams in Fig. 1 are denoted by the roman  
numbers from $I-VIII$. The diagrams $III$ and $IV$ represent nonresonant 
contribution (mechanism (i)). All the remaining diagrams, 
which proceed through the intermediate short-lived vector meson $V_0$ ($\rho$, $\omega$ and $\phi$),  
represent the {\it resonant contribution} (mechanism (ii)). The resonant amplitude is represented in the whole $q^2$ region by the Breit-Wigner vector meson propagator. 
($q$ is the sum of lepton and antilepton momenta). In the regions of 
$q^2$ far away from $m_{V0}^2$, the resonant amplitude is given therefore solely by the tail of the Breit-Wigner vector 
meson propagator. 
The square in each diagram of 
Fig. 1 denotes the weak transition due to the effective Lagrangian ${\cal 
L}_{LD}$ (\ref{weak}). This Lagrangian contains a product of two left handed 
quark currents $(\bar q_kq_l)^{\mu}$, each denoted by a dot on Fig. 1. The left handed currents will be expressed in terms of the relevant 
hadronic degrees of freedom: $D$, $D^*$, $P$ and $V$. In our notation the 
hadronic current $J_2$ in the 
diagram $II$, for example, creates $V$ meson, while 
the hadronic current $J_1$ annihilates $D$ and creates $V_0$ at the same time.

In the model we use, there is no contribution of $J/\Psi$ or other $\bar c c$ 
excited states. The contribution that would arise by the exchange of this mesons 
is effectively described by the diagram $III$ of Fig. 1, where $\bar c c$ 
exchange is ``hidden'' in the $DD^*\gamma$ coupling. The alternative approach of a direct $\bar 
c c$ exchange would require the knowledge of their couplings to photons over a 
wide region of $q^2$, of which one has only rudimentary knowledge \cite{DHT}.
\\
 
{\bf B. Short distance contributions due to $c\to ul^+l^-$}\\

In addition to long distance dynamics, the Cabibbo suppressed decays $D^0 \to 
\rho^0 l^+ l^-$, $D^0 \to \omega l^+ l^-$, 
$D^0 \to \phi l^+ l^-$, $D^+ \to \rho^+ l^+ l^-$ and 
$D_s^+ \to K^{*+} l^+ l^-$ can also be driven by the short distance $c\to 
ul^+l^-$ transition. The short distance part in $D\to Vl^+l^-$ decays will turn out in 
general to be much smaller than the long distance part. However, we shall find that in the 
case of $D^0 \to \rho^0 (\omega)l^+ l^-$ the short distance part is of the same 
order of magnitude as the nonresonant part of the long distance contribution. In this subsection we estimate the size of the short distance 
amplitudes, which is {\it nonresonant} in its nature.

The effective Lagrangian for FCNC transition $c \to u l^+ l^-$ arises from $WW$ exchange box 
diagrams and $Z$ and $\gamma^*$ penguin operators \cite{SCHWARTZ}. It has been 
obtained using the similar results for $s\to d l^+ l^-$ decay \cite{IL}
\begin{eqnarray}
\label{SD}
{\cal L}_{SD} & = & \frac{G_F}{{\sqrt 2}} \frac{e^2}{16 \pi^2 sin^2 \theta_W }
\nonumber\\
& \times& \sum_{i=d,s,b} V_i \biggl[~ {\bar u} \gamma_{\mu} (1- \gamma_5) c~ 
\biggl( A_i{\bar l} \gamma^{\mu} (1- \gamma_5) l +  
B_i{\bar l} \gamma^{\mu} (1+ \gamma_5) l \biggr) \nonumber\\
& - & 2i m_c sin^2 \theta_W F_2^i q^{\nu}~{\bar u} \sigma_{\mu \nu}  (1 
+\gamma_5) c ~
 {\bar l} \gamma^{\mu} (1- \gamma_5) l ~\biggr]~,
 \end{eqnarray} 
 where the Willson coefficients $A_i$, $B_i$ and $F_2^i$ are given in Appendix A and $V_i$ are the CKM 
coefficients, $V_i=V_{ci}^*V_{ui}$. The expression (\ref{SD}) does not contain 
the QCD 
corrections, which have not been studied for $c\to u l^+l^-$ decays so far.  
 When referring to these corrections for $c\to ul^+l^-$, one is reminded that in the case of 
$c\to u\gamma$ decay there is a huge QCD enhancement \cite{BGHP,GHMW}, which is  due to the 
following reason: The effect of QCD is that the Wilson coefficient $c_7(m_c)$, 
responsible for the magnetic penguin decay $c\to u\gamma$, obtains the admixture 
of the other Wilson coefficients evaluated at the scale $m_W$ in the leading 
order $c_i(m_W),~{i=1..10}$. Since $c_7(m_W)$ is extremely suppressed compared 
to some other Wilson coefficients $c_i(m_W)$, the resulting $c_7(m_c)$ is much 
bigger than $c_7(m_W)$ and a huge QCD enhancement for $c\to u\gamma$ occurs. In 
the $c\to u l^+l^-$ decay on the other hand, the responsible Wilson 
coefficients $A(m_W)$ and $B(m_W)$, which do not contribute for 
the real photons,  are not suppressed in the lowest order and one would not 
expect large QCD effects on them as one has learned from estimation of $s\to dl^+l^-$ 
\cite{SS}. The coefficient of the magnetic transition $F_2(m_W)$, which is 
proportional to Wilson coefficient $c_7(m_W)$, indeed acquires large QCD correction, however it is strongly suppressed in the lowest order.  Since we are only interested in the rough 
estimation of the short distance contribution in $D\to Vl^+l^-$ and the last 
term in (\ref{SD}) is much less important than the other two for such decays 
\cite{SS}, we neglect the last term altogether, using the approximation explained in Appendix A.  

The Lagrangian ${\cal L}_{SD}$ (\ref{SD}) then  gives the branching ratio for inclusive $c\to ul^+l^-$ process 
$${\Gamma(c\to ul^+l^-)\over \Gamma(D^0)}={G_F^2m_c^5\over 192\pi^3\Gamma(D^0)}\biggl({\alpha\over 
4\pi\sin^2\theta_W}\biggr)^2\bigl[|V_iA_i|^2+|V_iB_i|^2\bigr]=2.9~10^{-9}~.$$
To predict the exclusive amplitudes for $D\to Vl^+l^-$ induced by ${\cal L}_{SD}$ 
(\ref{SD}), we have to evaluate the 
matrix elements 
\begin{equation}
\label{sdfactor}
\langle V|{\bar u} \gamma_{\mu} (1- \gamma_5) c| D\rangle~.
\end{equation}
We shall do this by  again using the hybrid model, which is described in the next subsection.  The corresponding Feynman diagrams within this approach 
are given in Fig. 2. The squares in the diagrams denote the weak transition due 
to the short distance Lagrangian ${\cal L}_{SD}$ (\ref{SD}). This Lagrangian 
contains a product of a quark and lepton weak currents, each denoted by a dot in Fig. 2. We remark, that these diagrams have a long distance counterpart, given by diagrams $V$ and $VI$ of Fig. 1, which represent the long distance $c\to u\gamma$ transition \cite{FS,SS}  \\ 

{\bf  C. Theoretical framework: chiral Lagrangians, heavy quark limit and vector 
meson dominance}\\

Here we present the model, which we use to evaluate the matrix elements   
(\ref{factor}) and (\ref{sdfactor}) needed to predict $D\to Vl^+l^-$ amplitudes. The framework we use for our treatment is that of an effective Lagrangian, which embodies two important approximate symmetries of QCD, the infinite heavy quark $Q$ mass limit $(m_Q\to \infty)$ and the chiral limit for light quarks, namely $(m_u,~m_d,~m_s) \to 0$. This approach, which was developed during the last few years (\cite{BFO2}-\cite{cas} and additional references quoted in \cite{casrep}), has been used with a good measure of success to treat strong, electromagnetic and weak decays of $D$ and $B$ mesons. Obviously, an effective Lagrangian approach has also its weakness, as it involves a number of unknown coupling constants. Fortunately, the use of observed processes makes it possible to determine a good proportion of them, as detailed in this and the next subsection. Moreover, the use of form factors alleviates the limitations on the range in which the basic assumptions of the model hold to a good accuracy. The model and its various applications till now are well exposed in a recent review \cite{casrep}. In the present subsection, we describe those parts which are needed for our calculation.  
We introduce now the strong and electromagnetic interaction Lagrangians for the 
heavy (hadrons containing $c$ quark) and light (hadrons containing only light $u$, $d$ and $s$ quarks) sector and the relevant weak currents. At the end of the section 
we discuss the values of free parameters, that enter the Lagrangians and 
currents in our model. 
Our strong and electromagnetic Lagrangian \cite{BFO2}-\cite{casrep} is invariant under heavy quark spin ($SU(2)$), 
chiral ($SU(3)_L\times SU(3)_R$), Loretz, parity and $U(1)$ gauge transformation \cite{casrep}.  The light vector mesons are incorporated using the hidden symmetry approach \cite{casrep,bando}. We are aware that using HQET, which converges very slowly in the case of $c$ quark, presents a rather rough approximation. In spite of that, the HQET approach, which helps to reduce the number of free parameters, has been successfully applied in many $D$ decays (e.g. \cite{casrep} and references therein).

The light degrees of freedom are described by the 
3$\times$3 Hermitian matrices 

\begin{eqnarray}
\label{defpi}
\Pi = \pmatrix{
{\pi^0 \over \sqrt{2}} + {\eta_8 \over \sqrt{6}}+ {\eta_0 \over \sqrt{3}}& \pi^+ 
& K^+ \cr
\pi^- & {-\pi^0 \over \sqrt{2}} + {\eta_8 \over \sqrt{6}} + 
{\eta_0 \over \sqrt{3}} & K^0 \cr
K^- & {\bar K^0} & -{2 \eta_8 \over \sqrt{6}} + {\eta_0 \over \sqrt{3}} \cr}
\end{eqnarray}

\noindent
and

\begin{eqnarray}
\label{defrho}
\rho_\mu = \pmatrix{
{\rho^0_\mu + \omega_\mu \over \sqrt{2}} & \rho^+_\mu & K^{*+}_\mu \cr
\rho^-_\mu & {-\rho^0_\mu + \omega_\mu \over \sqrt{2}} & K^{*0}_\mu \cr
K^{*-}_\mu & {\bar K^{*0}}_\mu & \phi_\mu \cr}~, \quad F_{\mu \nu} ({\rho}) = \partial_\mu {\rho}_\nu -\partial_\nu {\rho}_\mu +
[{\rho}_\mu,{ \rho}_\nu]
\end{eqnarray}
\noindent
for the pseudoscalar and vector mesons, respectively. They are 
usually expressed through the combinations 
\begin{eqnarray}
\label{defu}
u & = & \exp  ( \frac{i \Pi}{f} )\;,
\end{eqnarray}
\noindent 
where $f\simeq f_{\pi}=132$ MeV is the pion pseudoscalar decay constant and 
\begin{eqnarray}
\label{defrhohat}
{\hat \rho}_\mu & = & i {{\tilde g}_V \over \sqrt{2}} \rho_\mu\;,
\end{eqnarray}
\noindent
where ${\tilde g}_V$ is fixed in the case of the exact flavor symmetry 
 to be the $VPP$ coupling ${\tilde g}_V=5.9$ \cite{bando}. 

The most general  strong Lagrangian for the light mesons in the leading order of 
chiral perturbation theory is
 \cite{bando}
\begin{eqnarray}
\label{deflight} 
{\cal L}_{light}^1  =  -{f^2 \over 2}
\{tr({\cal A}_\mu {\cal A}^\mu) +
a\, tr[({\cal V}_\mu - {\hat \rho}_\mu)^2]\}
+  {1 \over 2 {\tilde g}_V^2} tr[F_{\mu \nu}({\hat \rho})
F^{\mu \nu}({\hat \rho})]~,
\end{eqnarray}
where we have introduced two  currents
\begin{equation}
{\cal V}_{\mu} =  \frac{1}{2} (u^{\dag}
D_{\mu} u + u D_{\mu}u^{\dag})\quad {\rm and} \quad {\cal A}_{\mu}  =  
\frac{1}{2} (u^{\dag}D_{\mu} u - u D_{\mu}u^{\dag})~.
\end{equation}
Demanding the Lagrangian (\ref{deflight}) to be invariant under the local gauge 
transformation, corresponding to the electro-magnetic $U(1)$ transformation in 
QCD, we define the covariant derivatives as 
$$D_{\mu}u = (\partial_{\mu} + {\hat B}_{\mu} )u \qquad 
{\rm and}\qquad  D_{\mu}u^{\dag}= (\partial_{\mu} + {\hat B}_{\mu} )u^{\dag},$$  
with ${\hat B}_{\mu} = i e B_{\mu} Q$, 
$Q = diag (2/3,-1/3,-1/3)$ and $B_{\mu}$ being the photon field. 
 The constant $a$ (\ref{deflight}) is in principle a free parameter. We fix 
it to $a=2$ \cite{bando} assuming the exact vector meson dominance, where 
the pseudoscalars interact with the photon only through vector mesons. With this 
choice, the photon - vector meson interaction given by the second term of 
Lagrangian (\ref{deflight}) is  
\begin{eqnarray}
\label{VMD1} 
{\cal L}_{V\gamma} = -e {\tilde g}_V f^2 B_{\mu} (\rho^{0\mu} + \frac{1}{3} 
\omega^{\mu} - \frac{{\sqrt 2}}{3} \phi^{\mu})~.
\end{eqnarray}
Instead of using the exact $SU(3)$ symmetry values $\tilde g_V=5.9$ and 
$f=132~MeV$, we express the $V\gamma$ couplings in terms of the measurable quantities $g_{\rho}$, $g_{\omega}$ and $g_{\phi}$ defined by the matrix element of the corresponding vector current $J_V^{\mu}$
\begin{eqnarray} 
\label{defg}
\langle V(\epsilon_V,q)| J_V^{\mu} |0\rangle  = g_V(q^2)~\epsilon^{*\mu} (q) ~. 
\end{eqnarray}
In our calculation we use the values $g_V$ given in Table 2, which have been 
directly measured in the leptonic $V\to l^+l^-$ decays, and we make the assumption 
$g_{V}(q^2)=g_{V}(m_{V}^2) \equiv  g_{V}$.   
The photon - vector meson interaction Lagrangian (\ref{VMD1}) defined through 
the constants $g_V$ is
\begin{equation}
\label{VMD}
{\cal L}_{V\gamma} = -{e\over  \sqrt{2}}  (g_{\rho}\rho^{0\mu} + 
{g_{\omega}\over 3} \omega^{\mu} - {\sqrt{2}g_{\phi}\over 3}\phi^{\mu})~B_{\mu}.
\end{equation}
As far as the calculation of the amplitudes for the diagrams of Fig. 1 is 
concerned, the Lagrangian ${\cal L}_{light}$ (\ref{deflight}) provides also the 
$VVV$ vertex  given by the third term in 
(\ref{deflight}). The $VVV$ vertex is present in the diagram $VIII$ of Fig. 1, which 
describes the photon emission from the charged vector meson. 

We need also the $PVV$ vertex, which is present in the diagram $VII$ of Fig. 1. This interaction term can be generated only in the next-to-leading order of chiral perturbation theory as \cite{casrep} 
\begin{eqnarray}
\label{deflight2}
{\cal L}_{light}^2  =  -4 \frac{C_{VV\Pi}}{f} \epsilon
^{\mu \nu \alpha \beta}Tr (\partial_{\mu}
{\rho}_{\nu} \partial_{\alpha}{\rho}_{\beta} \Pi)~,
\end{eqnarray}
where $C_{VV\Pi}$ is free parameter.  

\vspace{2mm}

Both the heavy pseudoscalar and the heavy vector 
mesons are incorporated in a $4\times 4$ matrix 
\begin{eqnarray}
\label{defh}
H_a =  {1 + \!\!\not{\! v}\over 2} (P_{a\mu}^{*}
\gamma^{\mu} - P_{a} \gamma_{5})~,~~\quad {\bar H}_{a}  =  \gamma^{0} 
H_{a}^{\dag} \gamma^{0} =(P_{a\mu}^{* \dag} \gamma^{\mu} + P_{a}^{\dag} 
\gamma_{5})
{1 + \!\!\not{\! v}\over 2}~, 
\end{eqnarray}
\noindent
where $a=1,2,3$ is the $SU(3)_V$ index of the light 
flavors and $P_{a\mu}^*$, $P_{a}$ annihilate a 
spin $1$ and spin $0$ heavy meson $Q \bar{q}_a$ of 
velocity $v$, respectively. The strong and electromagnetic Lagrangian in the heavy sector have to provide us with the $DD\gamma$, $DD^*\gamma$ and $DD^*V$ vertices.  The first vertex describes the photon emission from the charged $D$ meson and is generated in the leading order of HQET (invariant under heavy quark symmetry and $U(1)$ gauge transformation with  minimal number of derivatives) as \cite{BFO1,casrep} 
\begin{eqnarray}
\label{defstrong}
{\cal L}_{heavy}^1  = i Tr [H_{a} v^{\mu} (\partial_{\mu} + {\cal V}_{\mu} - 
{2\over 3} i e  B_{\mu}) {\bar H}_{a}] 
\end{eqnarray} 
The  $DD^*\gamma$ and $DD^*V$ vertices can be generated only in the next-to-leading order of the heavy quark and chiral expansion and  are described by \cite{BFO1,casrep} 
\begin{eqnarray}
\label{defstrong2}
{\cal L}_{heavy}^2  =- {\lambda}^{\prime} Tr [H_{a}\sigma_{\mu \nu}
F^{\mu \nu} (B) {\bar H_{a}}]+i {\lambda} Tr [H_{a}\sigma_{\mu \nu}
F^{\mu \nu} (\hat \rho)_{ab} {\bar H_{b}}]~.
\end{eqnarray} 
The first term contributes to the diagram $III$  Fig. 1, while the second term contributes to the diagrams $I$ and $V$ of Fig. 1 and the diagram $I$ of Fig. 2.  The  $\lambda$ and $\lambda '$ are free 
parameters. 

In addition to the strong and electromagnetic interaction, we have to specify 
the weak one. The effective  weak Lagrangian responsible for the long distance contribution is given 
by ${\cal L}_{LD}$ (\ref{weak}) and for the short distance contribution by 
${\cal L}_{SD}$ (\ref{SD}). As we deal with the probabilities for the weak 
decays of hadrons, we rewrite the quark weak currents in ${\cal L}_{LD}$ 
(\ref{weak}) and ${\cal L}_{SD}$ (\ref{SD}) in terms 
of hadronic degrees of freedom.  The weak current 
${\bar q}_a\gamma^\mu(1-\gamma^5)c$ containing a $c$ quark and one light 
anti-quark 
$\bar q_a$ transforms under chiral $SU(3)_L\times SU(3)_R$ transformation as  $({\bar 
3}_L,1_R)$. At the hadronic level we impose the same chiral transformation and 
we require the current to be linear in the 
heavy meson fields $D^a$ and $D^{*a}_\mu$ \cite{BFO2,cas}
\begin{eqnarray}
\label{jqbig}
{J}_{a}^{\mu}& = &\frac{1}{2} i \alpha Tr [\gamma^{\mu}
(1 - \gamma_{5})H_{b}u_{ba}^{\dag}]\\
&+& \alpha_{1}  Tr [\gamma_{5} H_{b} ({\hat \rho}^{\mu}
- {\cal V}^{\mu})_{bc} u_{ca}^{\dag}]
+\alpha_{2} Tr[\gamma^{\mu}\gamma_{5} H_{b} v_{\alpha} 
({\hat \rho}^{\alpha}-{\cal V}^{\alpha})_{bc}u_{ca}^{\dag}]+...\;.\nonumber
\end{eqnarray}
The current (\ref{jqbig}) is the most general one in the leading $1/m_c$ order 
of HQET and next to leading order of chiral perturbation theory. The first term 
is connected to the definition of the 
heavy meson decay constant 
$\langle  D(p)| (\bar q_ac)^{\mu} |0\rangle = -if_Dp^{\mu}$, where 
$\alpha=f_D\sqrt{m_D}$. The second and third term contribute to the diagrams 
$II$ and $VI$ of Fig. 1 and diagram $II$ of Fig. 2, where $J_a^{\mu}$ 
(\ref{jqbig}) annihilates $D$ meson 
and creates $V$ or $V_0$ meson at the same time. The constants $\alpha_1$ and 
$\alpha_2$ are free parameters. \\

{\bf  D. The choice of the parameters}\\
 
We now turn to the values of the coupling constants $C_{VV\Pi}$, $\lambda$, 
$\lambda '$, $\alpha_1$ and $\alpha_2$, which we need in the evaluation of the 
amplitudes of diagrams on Figs. 1 and 2.

The coupling $C_{VV\Pi}$ can be  
determined in the case of the exact $SU(3)$ flavor symmetry 
following the hidden symmetry approach of \cite{bando} and 
is found to be $|C_{VV\Pi}| = 3{\tilde g}_V^2 /32 \pi^2 = 0.33$. 
Experimentally, it can be directly determined from the $V\to PV_0\to P\gamma$ 
decay rates. In the following we will use the average value of $C_{VV\Pi}$, 
obtained from the measurements of different $V\to PV_0\to P\gamma$ decays  {\bf 
$|C_{VV\Pi}|=0.31$} \cite{FS}, which is close to its SU(3) limit. 

We determine the three parameters $\lambda$, $\alpha_1$ and $\alpha_2$ using three 
values related to the helicity amplitudes $Br=0.048\pm0.004$, 
$\Gamma_L/\Gamma_T=1.23\pm0.13$ and $\Gamma_+/\Gamma_-=0.16\pm0.04$ for the 
process $D^+\to \bar K^{*0}l^+\nu_l$ \cite{BFO2}, taken from the average of data 
from different experiments \cite{PDG}. We get four sets of solutions for 
$\lambda$, $\alpha_1$ and $\alpha_2$ \cite{BFO2} and we choose the set, which 
gives the best fit with a number of the nonleptonic decays $D\to PV$, $D\to VV$ 
and $D\to PP$ \cite{BFOP}: 

\noindent
$ \lambda =-0.34\pm 0.07~,\quad \alpha_1 =-0.14\pm 0.01~, \quad {\rm and }\quad  
\alpha_2=-0.83\pm 0.4~.$

In order to gain information on $\lambda '$ we turn to an 
analysis of $D^{*0} \to D^0 \gamma$, 
$D^{*+}\to D^+ \gamma$ and $D_s^{*+} \to D^+_s \gamma$ decays. 
Experimentally, only the ratios 
$R_{\gamma}^0 = \Gamma (D^{*0} \to D^0 \gamma)/\Gamma 
(D^{*0} \to D^0 \pi^0)$ and 
$R_{\gamma}^+ = \Gamma (D^{*+} \to D^+ \gamma)/\Gamma 
(D^{*+} \to D^+ \pi^0)$ are known \cite{PDG}. 
Taking the $R_{\gamma}^0 = 0.616$ and $R_{\gamma}^+ = 0.036$ \cite{PDG}, 
we obtain two sets of solutions for $|\lambda^{\prime} /g|$ and 
$|\lambda/g|$, which gives two solutions for $|\lambda ^{\prime}/\lambda |$. The 
first is $|\lambda ^{\prime} /\lambda| = 0.77$ and the second is  
$|\lambda^{\prime} /\lambda |= 0.21$ \cite{FS}. 
Taking $\lambda =-0.34$ we get four possibilities for $ \lambda ^{\prime}=\pm 
0.26,\pm 0.071$, which all have to be considered.  
\\

{\bf III. THE AMPLITUDES AND BRANCHING RATIOS \\ FOR NINE $D\to Vl^+l^-$ DECAYS}\\

{\bf A. The amplitudes}\\

In this section we turn to the amplitudes and branching ratios for the nine $D\to 
Vl^+l^-$ decays. The interaction Lagrangians  (\ref{deflight}), (\ref{VMD}), (\ref{deflight2}),  (\ref{defstrong}), (\ref{defstrong2}) and 
the weak currents (\ref{defg}), (\ref{jqbig}) provide us with the vertices in the 
kinematical region, where the heavy quark and chiral symmetry are good (i.e. the 
velocity of the heavy mesons changes only slightly in the interaction and the 
energy of the light mesons is small). The problem is how to extrapolate the 
amplitudes to the rest of the kinematical region allowed in $D\to Vl^+l^-$ 
decays. We assume, that the 
vertices do not change significantly throughout the kinematical region, 
which is a reasonable assumption in $D$ decays. At the same time we use the full 
heavy meson propagators $1/(p^2-m^2)$ instead of the HQET propagators 
$1/(2mv\cdot k)$.  
We account for the short life time of the intermediate neutral 
vector meson $V_0$  by using the Breit Wigner form for the $V_0$ propagator
$$-i~{g_{\mu\nu}-{q_{\mu}q_{\nu}\over m_{V_0}^2}\over 
q^2-m_{V_0}^2+i\Gamma_{V_0}m_{V_0}}~,$$
where $\Gamma_{V_0}$ is the decay width of the $V_0$ meson and $q$ is its 
momentum.  Then, using the interaction Lagrangians (\ref{weak}), (\ref{SD}), 
(\ref{deflight}), (\ref{VMD}), (\ref{deflight2}),  (\ref{defstrong}), (\ref{defstrong2}) and the weak 
currents (\ref{defg}), (\ref{jqbig}), the calculation of the amplitudes for long 
distance diagrams on Fig. 1 and short distance diagrams on Fig. 2 is 
straightforward.  The calculated amplitudes for different diagrams have in 
general different Lorentz structure. It is convenient to treat the 
amplitudes of  similar Lorentz structure together. Then, the sum of the amplitudes 
within the model described above is given by the expression   
\begin{eqnarray}
\label{amplitude}
&{\cal A}&\!\!\!\![D(p)\to V(\epsilon_{(V)},p_{(V)})~l^+(p_+)~l^-(p_-)]  =  
 -\frac{G_F}{{\sqrt 2}}e^2 f_{Cab}~{1\over q^2}
\\ 
&\times&\!\!\! \epsilon_ {(V)\beta}~ \bar u(p_-)\gamma_{\nu}v(p_+)~ 
\bigl[\epsilon^{\mu \nu \alpha \beta} q_{\mu} 
p_{\alpha}A_{PC}+iA_{PV}^{\beta\nu}\bigr]\nonumber ~,
\end{eqnarray}
where $q=p_-+p_+$ is the momentum of the intermediate virtual photon and the 
corresponding Cabibbo factors $f_{Cab}$ are given in Table 1. $A_{PC}$ and 
$A_{PV}^{\beta\nu}$ correspond to parity conserving and parity violating 
amplitudes, respectively. They get contributions from different diagrams in 
Fig. 1 (long distance) and Fig. 2 (short distance). The short distance 
amplitudes  $A_{PC}$ and $A_{PV}^{\beta\nu}$ for  the Cabibbo suppressed decays 
are given in Appendix A. The long distance amplitudes  $A_{PC}$ and 
$A_{PV}^{\beta\nu}$ for  the Cabibbo  allowed, suppressed 
and doubly suppressed  decays are given in Appendix B. 

The decay width for $D\to Vl^+l^-$ is given by the square of the amplitude,  
summed over the polarizations of the three particles in the final state and 
integrated over the three body phase space
\begin{equation}
\label{gama}
\Gamma={1\over 2m_D(2\pi)^5}\sum_{polar.}\int |{\cal 
A}(p_{(V)},p_+,p_-)|^2 ~{d^3p_{(V)}\over 2p_{(V)}^0} {d^3p_{+}\over 2p_{+}^0} 
{d^3p_{-}\over 2p_{-}^0}~ \delta(p_{(V)}+p_++p_--p)~.
\end{equation}\\

{\bf B. Discussion of the results}\\

Firstly,  we present the results for the decays with the  muon final state $D\to 
V\mu^+\mu^-$ and we comment on the decays $D\to 
Ve^+e^-$ in the end.  
The branching ratios for the Cabibbo allowed, suppressed and doubly 
suppressed $D\to V\mu^+\mu^-$ 
decays are 
presented in Table 1. The last column presents the experimental upper bounds 
\cite{exp,PDG}.  The other columns present our theoretical predictions,
where the error bars are due to the uncertainty of 
the model parameters $\lambda '$ and $C_{VV\Pi}$, which can have any of the 
values $\lambda '=\pm 
0.07,\pm 0.26$ and $C_{VV\Pi}=\pm 0.31$. 
The total branching ratio $Br(total)$ 
containing long (Fig. 1) and short distance (Fig. 2) contributions 
is given in 
the fifth column. The third column presents the short distance part of the 
branching ratio $Br(SD)$ calculated from (\ref{SD}), which is present only in  the Cabibbo suppressed decays. The fourth column presents only the nonresonant part of the 
long distance contribution $Br(LD_{nonr})$. This part 
is bigger for the charged 
$D$ meson decays, where it is mainly due to the diagram $IV$ of Fig. 1. For the 
neutral $D$ meson decays, the diagram $IV$ vanishes and the remaining 
nonresonant diagram $III$ has smaller amplitude, which is proportional to 
$\lambda '$. The parameter  $\lambda '=\pm 
0.07,\pm 0.26$ has  large 
uncertainty and we are only able to quote the upper limit for 
$Br(LD_{nonr})$.  

Apart from the Cabibbo structure, the 
branching ratios depend mainly on whether the initial (and final) state is 
charged or neutral, with bigger branching ratio in the former case.  
We present also the  distributions  
$(1 /\Gamma_D ) d \Gamma (D\to V \mu^+\mu^-)/ d q^2$ as a function of $q^2$ ($q^2$ is invariant $\mu^+\mu^-$ mass)  
for the typical representatives of the Cabibbo allowed 
($D_s^+ \to \rho^+ \mu^+ \mu^-$   in Fig. 3 and   $D^0 \to \bar K^{*0} \mu^+ 
\mu^-$ in Fig. 4)
and suppressed ($D^0\to \rho^0\mu^+\mu^-$ in Fig. 5 and $D^+_s \to 
K^{*+}\mu^+\mu^-$ in Fig. 6), neutral
or charged $D$ meson decays. The short distance contribution (dot-dashed 
line) due to $c\to 
ul^+l^-$ transition is present only in the Cabibbo suppressed decays and it 
turns out to be much smaller than the long distance contribution. Concerning the long distance  
contribution, the resonant part is bigger than the nonresonant 
part (dashed line), except perhaps in the case of 
charged $D$ meson decays at the low $q^2$ (see Figs. 3-6). Note, that the nonresonant LD contribution  is generally 
smaller than the resonant LD contributions  even in the regions well outside the 
resonance peak at $q^2=m_{V_0}^2$. It is interesting to remark, that the short distance and nonresonant
long distance contributions are comparable for  Cabibbo suppressed neutral $D$ 
meson decays $D^0 \to \rho^0 l^+ l^-$   and $D^0 \to \omega l^+ l^-$.

In the decays with the electron final state $D\to Ve^+e^-$ the lowest 
kinematically allowed $q^2$ is $q_{min}^2=(2m_e)^2$, which is smaller than $q_{min}^2=(2m_{\mu})^2$ in the 
 $D\to 
V\mu^+\mu^-$ case.  In the 
region $q^2>(2m_{\mu})^2$ the electron rates are practically equal to the muon 
rates. In the region $q^2<(2m_{\mu})^2$, however, the rates for $D\to Ve^+e^-$  
are extremely enhanced due to the photon propagator $1/q^2$. However, the 
region down to $q_{min}^2=(2m_e)^2$ requires a more accurate treatment of the $q^2$ dependence when $q^2$ approaches to $0$ , which is 
beyond our scope here. We have calculated the $D\to Ve^+e^-$ branching ratios 
with the lower cut off $q^2=(2m_{\mu})^2$ and have obtained  values which 
are very close to the $D\to V\mu^+\mu^-$ branching ratios (the  $D\to 
V\mu^+\mu^-$ 
branching ratios are obtained integrating over the whole $q^2=[(2m_{\mu})^2,
(m_D-m_V)^2]$ region ).

The Cabibbo allowed decays 
$D^0 \to \bar K^{*0} \mu^+ \mu^-$  and  $D_s^+ \to \rho^+ \mu^+ \mu^-$ with 
the predicted branching ratios of the order $10^{-6}$ and $10^{-5}$, respectively,  
have the best probability for their early detection. Note that their 
branching ratios  are not far below the present experimental upper 
bound. 

In the Cabibbo suppressed decays, the short distance  contribution due to FCNC  
transition $c\to ul^+l^-$ has
branching ratio of order $Br(SD)\sim 10^{-10}$ and is therefore well masked 
by the long distance branching ratios of order $10^{-7}$. 
Obviously, to observe the FCNC transition $c\to u$ within the Standard Model, 
one must most likely look for other possibilities.
 Still, new physics could enhance the SD part to be of the same order as the LD part or bigger \cite{hew,castro,BABU}. 
In this case the branching ratios well above $10^{-7}$ for  
Cabibbo suppressed decays $D\to V\mu^+\mu^-$ would signal new physics. As the present experimental upper 
bound is much higher, these decays still contain a large discovery window.  \\   

{\bf IV. SUMMARY}\\

We have calculated the long and short distance contributions for nine $D\to 
Vl^+l^-$ decays within the Standard Model. The short distance 
contribution is present only 
in the Cabibbo suppressed decays and is due to the flavour changing neutral 
transition $c\to ul^+l^-$. The long distance contribution is composed of the 
resonant part, which arises from the intermediate light vector meson $V_0$ 
exchange ($D\to VV_0\to V\gamma \to Vl^+l^-$), and the nonresonant part, which 
arises from the direct photon emission ($D\to  V\gamma \to Vl^+l^-$). 
The branching ratios are calculated using an effective Lagrangian, which combines heavy quark symmetry and chiral 
perturbation theory, and 
are given in Table 1.
The most frequent 
decays   are the Cabibbo allowed decays, which are expected at the rates, that are not much lower than the present experimental upper limit: $D_s^+\to 
\rho^+\mu^+\mu^-$ is expected at the branching ratio of approximately $3\cdot 10^{-5}$, while $D^0\to\bar K^{*0}\mu^+\mu^-$ is expected at $1.7\cdot 10^{-6}$. 
The Cabibbo suppressed 
decays on the other hand, are typically expected at $[3-7]\cdot 10^{-7}$ range 
for $D^0\to \rho^0(\omega)\mu^+\mu^-$ and $D_s^+\to K^{*+}\mu^+\mu^-$ decays and 
in the $10^{-6}$ range for $D^+\to\rho^+\mu^+\mu^-$ decay.  
Accordingly, 
branching ratios well above $10^{-7}$ for  Cabibbo suppressed decays could 
signal new physics. In all the Cabibbo suppressed decays the short distance 
contribution is  well masked in the Standard Model by the resonant long distance 
contribution. In the case of $D^0\to \rho^0(\omega)l^+l^-$ decays, however, the 
short distance contribution is of comparable size as the nonresonant long 
distance part.\\ 

\begin{table}[h]
\begin{center}
\begin{tabular}{|c||c||c|c||c||c|}
\hline
 $D\to V \mu^+\mu^-$ & $f_{Cab}$  & $Br(SD)$ & $Br(LD_{nonr})$ & $Br(total)$ & $Br(exp)$ \\
\hline \hline
  $ D^0 \to {\bar K}^{*0} \mu^+\mu^-$ & $a_2c^2$ & $0$ &$\leq 
1.9~10^{-8}$ &  $[1.6-1.9]~ 10^{-6}$ & $<~1.18~10^{-3}$  \\
\hline
  $ D_s^+ \to \rho^+ \mu^+\mu^-$  &$ a_1c^2$ & $0$ & $ 
4.0~10^{-6}$ & $[3.0-3.3] ~10^{-5}$ &   \\
\hline
\hline
 $ D^0 \to \rho^{0}\mu^+\mu^-$ &$ -a_2sc$ & $9.7~ 10^{-10}$ & $\leq 
4.8~10^{-10}$ & $[3.5-4.7]~ 10^{-7}$ & $<~2.3~10^{-4}$\\
\hline
 $ D^0 \to \omega \mu^+\mu^-$ &$-a_2sc$ & $9.1~ 10^{-10}$ & $\leq 
3.7~10^{-10}$ & $[3.3-4.5]~ 10^{-7}$  & $<~8.3~10^{-4}$\\
\hline
 $ D^0 \to \phi \mu^+\mu^-$ &$ a_2sc$ & $0$ & $\leq 
1.1~10^{-9}$ & $[6.5-9.0]~ 10^{-8}$ & $<~4.1~10^{-4}$\\
\hline
 $ D^+ \to \rho^+ \mu^+\mu^-$ &$ -a_1sc$ & $4.8~ 10^{-9}$ & 
$2.7~10^{-7}$ & $[1.5-1.8]~ 10^{-6}$ & $<~5.6~10^{-4}$ \\
\hline
 $ D_s^+ \to K^{*+ }\mu^+\mu^-$ &$ a_1sc$ & $1.6~10^{-9}$ & $1.5~ 
10^{-7}$ & $[5.0-7.0]~ 10^{-7}$ & $<~1.4~10^{-3}$\\
\hline
\hline 
 $ D^+ \to K^{*+} \mu^+\mu^-$ &$ -a_1s^2$ & $0$ & $1.0 
~10^{-8}$ & $[3.1-3.7]~ 10^{-8}$ & $<~8.5~10^{-4}$\\
\hline
 $ D^0 \to K^{*0} \mu^+\mu^-$ &$ -a_2s^2$ & $0$ & $\leq 
5.0~10^{-11}$ & $[4.4-5.1]~ 10^{-9}$ & \\
\hline
\end{tabular}
\caption{ The branching ratios for the Cabibbo allowed, suppressed and doubly 
suppressed $D\to V \mu^+\mu^-$ decays.
The last column presents 
the experimental upper bounds \cite{exp,PDG}, while 
the other columns present our theoretical predictions. 
The total branching ratio 
$Br(total)$ containing long (Fig. 1) and short distance (Fig. 2) contributions 
is given in the fifth column.  The third column presents only the short distance 
part of the branching ratio $Br(SD)$. The fourth column presents only the 
nonresonant part of the long distance contribution $Br(LD_{nonr})$. 
The error bars in the Table are due to the uncertainty of the model parameters  
expressed by the possibilities $\lambda '=\pm 
0.07,\pm 0.26$ and $C_{VV\Pi}=\pm 0.31$. The branching ratios for $D\to Ve^+e^-$ 
obtained with the lower cut off $q^2=(2m_{\mu})^2$ ($q^2$ the invariant $\mu^+ 
\mu^-$ mass) are almost exactly the same as  the branching ratios 
for $D\to V\mu^+\mu^-$ 
given in this Table. The second column gives the corresponding Cabibbo factors $f_{Cab}$ in 
terms of the Cabibbo angle $c=\cos \theta_C$ and $s=\sin\theta_C$.   }
\end{center}
\end{table}

\newpage

{\bf APPENDIX A: The short distance amplitudes}\\

In this Appendix we list the values of the Willson coefficients $A_i$, $B_i$ and $F_2^i$ 
in the short distance Lagrangian ${\cal L}_{SD}$ (\ref{SD}) and give the 
resulting short distance amplitudes $A_{PC}$ and $A_{PV}^{\beta\nu}$ from 
(\ref{amplitude}). 

The coefficients $A_i$, $B_i$ and $F_2^i$ have been obtained in the leading 
order by Inami and Lim \cite{IL} and following the notation of \cite{SCHWARTZ} 
one has 
\begin{eqnarray}
\label{ab}
A_i&=&C_i^{box}+C_i^Z-\sin ^2\theta_W(F_1^i+C_i^Z)~,\nonumber\\
B_i&=&-\sin ^2\theta_W(F_1^i+C_i^Z)~, 
\end{eqnarray}
where $C_i^{box}$, $C_i^Z$, $F_1^i$ and $F_2^i$ are kinematic factors, which 
depend on the $i$th-quark mass through $x_i=m_i^2/m_W^2$
\begin{eqnarray}
\label{willson}
C_i^{box}&=&{3\over 8} \biggl[-{1\over x_i-1}+{x_i\ln x_i \over 
(x_i-1)^2}\biggr]-\gamma (\xi,x_i)
\\
C_i^Z&=&{x_i\over 4}-{3\over 8}{1\over x_i-1}+{3\over 8}{2x_i^2-x_i\over 
(x_i-1)^2}\ln x_i +\gamma (\xi,x_i)\nonumber\\
F_1^i&=&Q\Biggl(\biggl[{1\over 12}{1\over x_i-1}+{13\over 12}{1\over 
(x_i-1)^2}-{1\over 2}{1\over (x_i-1)^3}\biggr]x_i\nonumber\\
&+&\biggl[{2\over 3}{1\over x_i-1}+\biggl({2\over 3}{1\over (x_i-1)^2}-{5\over 
6}{1\over (x_i-1)^3}+{1\over 2}{1\over (x_i-1)^4}\biggr)x_i\biggr]\ln 
x_i\Biggr)\nonumber\\
&-&\biggl[{7\over 3}{1\over x_i-1}+{13\over 12}{1\over (x_i-1)^2}-{1\over 
2}{1\over (x_i-1)^3}\biggr]x_i\nonumber\\
&-&\biggl[{1\over 6}{1\over x_i-1}-{35\over 12}{1\over (x_i-1)^2}-{5\over 
6}{1\over (x_i-1)^3}+{1\over 2}{1\over (x_i-1)^4}\biggr]x_i\ln x_i-2\gamma
(\xi,x_i)\nonumber\\
F_2^i&=&-Q\Biggl(\biggl[-{1\over 4}{1\over x_i-1}+{3\over 4}{1\over 
(x_i-1)^2}+{3\over 2}{1\over (x_i-1)^3}\biggr]x_i-{3\over 2}{x_i^2\ln x_i\over 
(x_i-1)^4}\Biggr)\nonumber\\
&+&\biggl[{1\over 2}{1\over x_i-1}+{9\over 4}{1\over (x_i-1)^2} +{3\over 
2}{1\over (x_i-1)^3}\biggr]x_i-{3\over 2}{x_i^3\ln x_i\over (x_i-1)^4}
\end{eqnarray}
The summation $i$ in ${\cal L}_{SD}$ (\ref{SD}) runs over down-like quarks ($d$, 
$s$ and $b$) to which charm can couple, while $Q=-1/3$ is the corresponding 
charge of the intermediate quarks (we note that $F_1$ and  $F_2$  have been 
calculated in \cite{SCHWARTZ} using the wrong charge $Q=2/3$).  The gauge dependent term $\gamma 
(\xi ,x_i)$ \cite{IL} cancels out in the combinations $A_i$, $B_i$ and $F_2^i$ 
(\ref{ab}). Since the ratios $x_d$, $x_s$ and $x_b$ are of orders $10^{-8}$, 
$10^{-6}$ and $10^{-3}$ respectively, the terms proportional to the powers of 
$x_i$ can be safely neglected in (\ref{willson}). With this approximation 
$C_i^{box}=C_i^Z=-3/8$, $F_1^i=-2 \ln x_i/(9x_i-9)$ and $F_2^i$ vanishes. In 
this limit the GIM cancelation occurs and we obtain 
\begin{equation}
\sum V_iA_i=\sum V_iB_i\equiv A_{SD}=-0.065
\end{equation}
 Consequently, the short distance Lagrangian (\ref{SD}) 
effectively contains only the vector lepton current $\bar l\gamma^{\mu}l$ but 
not the axial vector $\bar l\gamma^{\mu}\gamma_5 l$ one. 

\vspace{0.2cm}

The short distance $c\to ul^+l^-$ transition contributes only to the Cabibbo 
suppressed decays. 
Here we give the short distance contributions  for the parity conserving 
$A_{PC}$ and parity violating  $A^{\beta\nu}_{PV}$ amplitudes, which are needed 
to calculate the Cabibbo suppressed amplitudes ${\cal A}[D(p)\to 
V(\epsilon_{(V)},p_{(V)})~l^+(p_+)~l^-(p_-)]$, (\ref{amplitude}). Within the 
model used, these amplitudes are given by the diagrams on Fig. 2: 
\begin{eqnarray}
A_{PC}(D^0\!\to\! \rho^0 l^+l^-)&\!\!\!=\!\!\!&A_{PC}(D^0\!\to\! \omega 
l^+l^-)=A_{PC}(D^+\!\to\! \rho^+ l^+l^-)/\sqrt{2}\nonumber\\
=A_{PC}(D_s^+\!\to\! K^{*+} l^+l^-)/\sqrt{2}&\!\!\!=\!\!\!&
4{f_{D^*}\lambda \tilde g_V \over f_{Cab}} \sqrt{{m_{D*}\over 
m_{D}}}{m_{D*}\over q^2-m_{D*}^2}{A_{SD}~q^2\over 16\pi^2\sin^2\theta_W}~,
~\nonumber\\
\nonumber\\
A_{PV}^{\beta\nu}(D^0\!\to\! \rho^0 
l^+l^-)&\!\!\!=\!\!\!&A_{PV}^{\beta\nu}(D^0\!\to\! \omega 
l^+l^-)=A_{PV}^{\beta\nu}(D^+\!\to\! \rho^+ l^+l^-)/\sqrt{2}\nonumber\\
=A_{PV}^{\beta\nu}(D_s^+\!\to\! K^{*+} l^+l^-)/\sqrt{2}&\!\!\!=\!\!\!&-2{\tilde 
g_V\sqrt{m_D}\over f_{Cab}}~\biggl[\alpha_1 
g^{\beta\nu}-\alpha_2{q^{\beta}p^{\nu}\over m_D^2}\biggr]{A_{SD}~q^2\over 
16\pi^2\sin^2\theta_W}~,
~\nonumber\\
\nonumber\\
A_{PC}^{\beta\nu}(D^0\!\to\! \phi^0 
l^+l^-)&\!\!\!=\!\!\!&A_{PV}^{\beta\nu}(D^0\!\to\! \phi l^+l^-)=0~.
\end{eqnarray}
The last equation is a result of Eq. (\ref{sdfactor}) and the quark content of 
the $\phi$ meson. 
The relevant constants are presented in Appendix B.\\

{\bf APPENDIX B: The long distance amplitudes}\\

In this Appendix we give the expressions for the parity conserving $A_{PC}$ and 
parity violating amplitudes $A^{\beta\nu}_{PV}$, which are needed to calculate 
the amplitudes ${\cal A}[D(p)\to 
V(\epsilon_{(V)},p_{(V)})~l^+(p_+)~l^-(p_-)]$ (\ref{amplitude}) for nine $D\to V l^+l^-$ decays. The following amplitudes 
$A_{PC}$ and  $A^{\beta\nu}_{PV}$ contain the long distance resonant and 
nonresonant contributions coming from the diagrams on Fig. 1. The coefficients 
and constants needed for the evaluation of the amplitudes will be given bellow:
\begin{eqnarray}
\label{aa}
A_{PC}(D^0\!\to\!\bar K^{*0}l^+l^-)\!\!\!& = &\!\!\!4J^{D^0}g_{K*} 
f_{D*}\sqrt{{m_{D*}\over m_{D}}}{m_{D*}\over m_{K*}^2-m_{D*}^2}-2K^{\bar 
K^{*0}}C_{VV\Pi}f_Dm_D^2
~,\nonumber\\
\nonumber\\
A^{\beta\nu}_{PV}(D^0\!\to\! \bar 
K^{*0}l^+l^-)\!\!\!&=&\!\!\!M^{D^0}g_{K*}\sqrt{m_D}~\biggl[\alpha_1 
g^{\beta\nu}-\bigl(\alpha_1-\alpha_2{q\cdot p\over 
m_D^2}\bigr){q^{\beta}q^{\nu}\over m_{V_0}^2}-\alpha_2{q^{\beta}p^{\nu}\over 
m_D^2}\biggr]~,
~\nonumber\\
\nonumber\\
A_{PC}(D_s^+\!\to\! \rho^+l^+l^-)\!\!\!& = &\!\!\!4J^{D_s^+}g_{\rho} 
f_{D_s*}\sqrt{{m_{D_s*}\over m_{D_s}}}{m_{D_s*}\over 
m_{\rho}^2-m_{D_s*}^2}-2K^{\rho^+}C_{VV\Pi}f_{D_s}m_{D_s}^2~,
\nonumber\\
\nonumber\\
A^{\beta\nu}_{PV}(D_s^+\!\to\! 
\rho^+l^+l^-)\!\!\!&=&\!\!\!2f_{D_s}g_{\rho}\biggl[{q^{\beta}p^{\nu}\over 
m_{D_s}^2-m_{\rho}^2}-L^{\rho^+}{g^{\beta\nu}\over 2}(q^2-m_{\rho}^2)\biggr]
\nonumber\\
\nonumber\\
\!\!\!&+&\!\!\!M^{D_s^+}g_{\rho}\sqrt{m_{D_s}}~\biggl[\alpha_1 
g^{\beta\nu}-\bigl(\alpha_1-\alpha_2{q\cdot p\over 
m_{D_s}^2}\bigr){q^{\beta}q^{\nu}\over m_{V_0}^2}-\alpha_2{q^{\beta}p^{\nu}\over 
m_{D_s}^2}\biggr]~,
~\nonumber\\
\nonumber\\
A_{PC}(D^0\!\to\! \rho^0 l^+l^-)\!\!\!& = &\!\!\!-4{J^{D^0}\over 
\sqrt{2}}g_{\rho} 
f_{D*}\sqrt{{m_{D*}\over m_{D}}}{m_{D*}\over 
m_{\rho}^2-m_{D*}^2}-2K^{\rho^0}C_{VV\Pi}f_Dm_D^2
\nonumber\\
\nonumber\\
\!\!\!& - &\!\!\!Nf_{D^*}\lambda \tilde g_V {a_2\sin\theta_C\cos\theta_C\over 
f_{Cab}}\sqrt{{m_{D*}\over m_{D}}}{m_{D*}\over q^2-m_{D*}^2}~,
\nonumber\\
\nonumber\\
A^{\beta\nu}_{PV}(D^0\!\to\! \rho^0 
l^+l^-)\!\!\!&=&\!\!\!M^{D^0}g_{\rho}\sqrt{m_D}~\biggl[\alpha_1 
g^{\beta\nu}-\bigl(\alpha_1-\alpha_2{q\cdot p\over 
m_D^2}\bigr){q^{\beta}q^{\nu}\over m_{V_0}^2}-\alpha_2{q^{\beta}p^{\nu}\over 
m_D^2}\biggr]
\nonumber\\
\nonumber\\
\!\!\!&+&\!\!\!{N\over 2}~\tilde g_V\sqrt{m_D}~{a_2\sin\theta_C\cos\theta_C\over 
f_{Cab}}~\biggl[\alpha_1 g^{\beta\nu}-\alpha_2{q^{\beta}p^{\nu}\over 
m_D^2}\biggr]~,
~\nonumber\\
\nonumber\\
A_{PC}(D^0\!\to\! \omega l^+l^-)\!\!\!& = &\!\!\!4{J^{D^0}\over 
\sqrt{2}}g_{\omega} 
f_{D*}\sqrt{{m_{D*}\over m_{D}}}{m_{D*}\over 
m_{\omega}^2-m_{D*}^2}-2K^{\omega}C_{VV\Pi}f_Dm_D^2
\nonumber\\
\nonumber\\
\!\!\!& - &\!\!\!Nf_{D^*}\lambda \tilde g_V {a_2\sin\theta_C\cos\theta_C\over 
f_{Cab}}\sqrt{{m_{D*}\over m_{D}}}{m_{D*}\over q^2-m_{D*}^2}~,
\nonumber\\
\nonumber\\
A^{\beta\nu}_{PV}(D^0\!\to\! \omega 
l^+l^-)\!\!\!&=&\!\!\!M^{D^0}g_{\omega}\sqrt{m_D}~\biggl[\alpha_1 
g^{\beta\nu}-\bigl(\alpha_1-\alpha_2{q\cdot p\over 
m_D^2}\bigr){q^{\beta}q^{\nu}\over m_{V_0}^2}-\alpha_2{q^{\beta}p^{\nu}\over 
m_D^2}\biggr]
\nonumber\\
\nonumber\\
\!\!\!&+&\!\!\!{N\over 2}~\tilde g_V\sqrt{m_D}~{a_2\sin\theta_C\cos\theta_C\over 
f_{Cab}}~\biggl[\alpha_1 g^{\beta\nu}-\alpha_2{q^{\beta}p^{\nu}\over 
m_D^2}\biggr]~,
~\nonumber\\
\nonumber\\
A_{PC}(D^0\!\to\! \phi l^+l^-)\!\!\!& = &\!\!\!4J^{D^0}g_{\phi} 
f_{D*}\sqrt{{m_{D*}\over m_{D}}}{m_{D*}\over 
m_{\phi}^2-m_{D*}^2}-2K^{\phi}C_{VV\Pi}f_Dm_D^2~,
\nonumber\\
\nonumber\\
A^{\beta\nu}_{PV}(D^0\!\to\! \phi 
l^+l^-)\!\!\!&=&\!\!\!M^{D^0}g_{\phi}\sqrt{m_D}~\biggl[\alpha_1 
g^{\beta\nu}-\bigl(\alpha_1-\alpha_2{q\cdot p\over 
m_D^2}\bigr){q^{\beta}q^{\nu}\over m_{V_0}^2}-\alpha_2{q^{\beta}p^{\nu}\over 
m_D^2}\biggr]~,
~\nonumber\\
\nonumber\\
A_{PC}(D^+\!\to\! \rho^+l^+l^-)\!\!\!& = &\!\!\!4J^{D^+}g_{\rho} 
f_{D*}\sqrt{{m_{D*}\over m_{D}}}{m_{D*}\over 
m_{\rho}^2-m_{D*}^2}-2K^{\rho^+}C_{VV\Pi}f_Dm_D^2
\nonumber\\
\nonumber\\
\!\!\!& - &\!\!\!\sqrt{2} Nf_{D^*}\lambda \tilde g_V 
{a_2\sin\theta_C\cos\theta_C\over f_{Cab}}\sqrt{{m_{D*}\over 
m_{D}}}{m_{D*}\over q^2-m_{D*}^2}~,
\nonumber\\
\nonumber\\
A^{\beta\nu}_{PV}(D^+\!\to\! 
\rho^+l^+l^-)\!\!\!&=&\!\!\!2f_Dg_{\rho}\biggl[{q^{\beta}p^{\nu}\over 
m_D^2-m_{\rho}^2}-L^{\rho^+}{g^{\beta\nu}\over 2}(q^2-m_{\rho}^2)\biggr]
\nonumber\\
\nonumber\\
\!\!\!&+&\!\!\!M^{D^+}g_{\rho}\sqrt{m_D}~\biggl[\alpha_1 
g^{\beta\nu}-\bigl(\alpha_1-\alpha_2{q\cdot p\over 
m_D^2}\bigr){q^{\beta}q^{\nu}\over m_{V_0}^2}-\alpha_2{q^{\beta}p^{\nu}\over 
m_D^2}\biggr]
\nonumber\\
\nonumber\\
\!\!\!&+&\!\!\!{N\over \sqrt{2}}~\tilde 
g_V\sqrt{m_D}~{a_2\sin\theta_C\cos\theta_C\over f_{Cab}}~\biggl[\alpha_1 
g^{\beta\nu}-\alpha_2{q^{\beta}p^{\nu}\over m_D^2}\biggr]~,
~\nonumber\\
\nonumber\\
A_{PC}(D_s^+\!\to\! K^{*+}l^+l^-)\!\!\!& = &\!\!\!4J^{D_s^+}g_{K*} 
f_{D_s*}\sqrt{{m_{D_s*}\over m_{D_s}}}{m_{D_s*}\over 
m_{K*}^2-m_{D_s*}^2}-2K^{K^{*+}}C_{VV\Pi}f_{D_s}m_{D_s}^2
\nonumber\\
\nonumber\\
\!\!\!& - &\!\!\!\sqrt{2} Nf_{D^*}\lambda \tilde g_{V} 
{a_2\sin\theta_C\cos\theta_C\over f_{Cab}}\sqrt{{m_{D*}\over 
m_{D_s}}}{m_{D*}\over q^2-m_{D*}^2}~,
\nonumber\\
\nonumber\\
A^{\beta\nu}_{PV}(D_s^+\!\to\! 
K^{*+}l^+l^-)\!\!\!&=&\!\!\!2f_{D_s}g_{K*}\biggl[{q^{\beta}p^{\nu}\over 
m_{D_s}^2-m_{K*}^2}-L^{K^{*+}}{g^{\beta\nu}\over 2}(q^2-m_{K*}^2)\biggr]
\nonumber\\
\nonumber\\
\!\!\!&+&\!\!\!M^{D_s^+}g_{K*}\sqrt{m_{D_s}}~\biggl[\alpha_1 
g^{\beta\nu}-\bigl(\alpha_1-\alpha_2{q\cdot p\over 
m_{D_s}^2}\bigr){q^{\beta}q^{\nu}\over m_{V_0}^2}-\alpha_2{q^{\beta}p^{\nu}\over 
m_{D_s}^2}\biggr]
\nonumber\\
\nonumber\\
\!\!\!&+&\!\!\!{N\over \sqrt{2}}~\tilde 
g_{V}\sqrt{m_{D_s}}~{a_2\sin\theta_C\cos\theta_C\over f_{Cab}}~\biggl[\alpha_1 
g^{\beta\nu}-\alpha_2{q^{\beta}p^{\nu}\over m_{D_s}^2}\biggr]~,
~\nonumber\\
\nonumber\\
A_{PC}(D^+\!\to\! K^{*+}l^+l^-)\!\!\!& = &\!\!\!4J^{D^+}g_{K*} 
f_{D*}\sqrt{{m_{D*}\over m_{D}}}{m_{D*}\over 
m_{K*}^2-m_{D*}^2}-2K^{K^{*+}}C_{VV\Pi}f_Dm_D^2~,
\nonumber\\
\nonumber\\
A^{\beta\nu}_{PV}(D^+\!\to\! 
K^{*+}l^+l^-)\!\!\!&=&\!\!\!2f_Dg_{K*}\biggl[{q^{\beta}p^{\nu}\over 
m_D^2-m_{K*}^2}-L^{K^{*+}}{g^{\beta\nu}\over 2}(q^2-m_{K*}^2)\biggr]
\nonumber\\
\nonumber\\
\!\!\!&+&\!\!\!M^{D^+}g_{K*}\sqrt{m_D}~\biggl[\alpha_1 
g^{\beta\nu}-\bigl(\alpha_1-\alpha_2{q\cdot p\over 
m_D^2}\bigr){q^{\beta}q^{\nu}\over m_{V_0}^2}-\alpha_2{q^{\beta}p^{\nu}\over 
m_D^2}\biggr]~,
~\nonumber\\
\nonumber\\
A_{PC}(D^0\!\to\!  K^{*0}l^+l^-)\!\!\!& = &\!\!\!4J^{D^0}g_{K*} 
f_{D*}\sqrt{{m_{D*}\over m_{D}}}{m_{D*}\over m_{K*}^2-m_{D*}^2}-2K^{\bar 
K^{*0}}C_{VV\Pi}f_Dm_D^2~,
\nonumber\\
\nonumber\\
A^{\beta\nu}_{PV}(D^0\!\to\! 
K^{*0}l^+l^-)\!\!\!&=&\!\!\!M^{D^0}g_{K*}\sqrt{m_D}~\biggl[\alpha_1 
g^{\beta\nu}-\bigl(\alpha_1-\alpha_2{q\cdot p\over 
m_D^2}\bigr){q^{\beta}q^{\nu}\over m_{V_0}^2}-\alpha_2{q^{\beta}p^{\nu}\over 
m_D^2}\biggr]~,
~\nonumber\\
\nonumber\\
\end{eqnarray}
Here $q=p_-+p_+$ and $m_{V_0}$ can be approximately taken as the average of the 
$\phi$, $\omega$ and $\rho$ masses. The coefficients $J^D$, $K^V$, $L^V$, $M^D$ 
and $N$ are expressed as
\begin{eqnarray}
\label{jklm}
J^{D^0}\!\!\!&=&\!\!\!\lambda '-{\lambda\tilde g_V\over 
2\sqrt{2}}\biggl[{g_{\rho}\over 
q^2-m_{\rho}^2+i\Gamma_{\rho}m_{\rho}}+{g_{\omega}\over 
3(q^2-m_{\omega}^2+i\Gamma_{\omega}m_{\omega})}\biggr]~,
\\
J^{D^+}\!\!\!&=&\!\!\!\lambda '-{\lambda\tilde g_V\over 
2\sqrt{2}}\biggl[-{g_{\rho}\over 
q^2-m_{\rho}^2+i\Gamma_{\rho}m_{\rho}}+{g_{\omega}\over 
3(q^2-m_{\omega}^2+i\Gamma_{\omega}m_{\omega})}\biggr]~,
\nonumber\\
J^{D_s^+}\!\!\!&=&\!\!\!\lambda '+{\lambda\tilde g_V\over 
2\sqrt{2}}~{2g_{\phi}\over 3(q^2-m_{\phi}^2+i\Gamma_{\phi}m_{\phi})}~,
\nonumber\\
K^{\bar K^{*0}}\!\!\!&=&\!\!\!\biggl[{g_{\rho}\over 
q^2-m_{\rho}^2+i\Gamma_{\rho}m_{\rho}}-{g_{\omega}\over 
3(q^2-m_{\omega}^2+i\Gamma_{\omega}m_{\omega})}+{2g_{\phi}\over 
3(q^2-m_{\phi}^2+i\Gamma_{\phi}m_{\phi})}\biggr]{1\over m_D^2-m_K^2}~,
\nonumber\\
K^{K^{*+}}\!\!\!&=&\!\!\!\biggl[-{g_{\rho}\over 
q^2-m_{\rho}^2+i\Gamma_{\rho}m_{\rho}}-{g_{\omega}\over 
3(q^2-m_{\omega}^2+i\Gamma_{\omega}m_{\omega})}+{2g_{\phi}\over 
3(q^2-m_{\phi}^2+i\Gamma_{\phi}m_{\phi})}\biggr]{1\over m_D^2-m_K^2}~,
\nonumber\\
K^{\rho^+}\!\!\!&=&\!\!\!-{2g_{\omega}\over 
3(q^2-m_{\omega}^2+i\Gamma_{\omega}m_{\omega})}{1\over m_D^2-m_{\pi}^2}~,
\nonumber\\
K^{\rho^0}\!\!\!&=&\!\!\!-2\sqrt{2}{g_{\rho}\over 
q^2-m_{\rho}^2+i\Gamma_{\rho}m_{\rho}}\biggl[{f_{1mix}(f_{1mix}-f_{2mix})\over 
m_D^2-m_{\eta ^2}}+{f_{1mix}^{'}(f_{1mix}^{'}-f_{2mix}^{'})\over m_D^2-m_{\eta 
'}}\biggr]~,
\nonumber\\
\!\!\!&+&\!\!\!{\sqrt{2}\over 3}{g_{\omega}\over 
(q^2-m_{\omega}^2+i\Gamma_{\omega}m_{\omega})}{1\over m_D^2-m_{\pi}^2}~,
\nonumber\\
K^{\omega}\!\!\!&=&\!\!\!-2\sqrt{2}{g_{\omega}\over 
q^2-m_{\omega}^2+i\Gamma_{\omega}m_{\omega}}\biggl[{f_{1mix}(f_{1mix}-f_{2mix})\
over m_D^2-m_{\eta ^2}}+{f_{1mix}^{'}(f_{1mix}^{'}-f_{2mix}^{'})\over 
m_D^2-m_{\eta '}}\biggr]~,
\nonumber\\
\!\!\!&+&\!\!\!\sqrt{2}{g_{\rho}\over 
q^2-m_{\rho}^2+i\Gamma_{\rho}m_{\rho}}{1\over m_D^2-m_{\pi}^2}~,
\nonumber\\
K^{\phi}\!\!\!&=&\!\!\!{4\over 3}{g_{\phi}\over( 
q^2-m_{\phi}^2+i\Gamma_{\phi}m_{\phi})}\biggl[{f_{1mix}(f_{1mix}-f_{2mix})\over 
m_D^2-m_{\eta ^2}}+{f_{1mix}^{'}(f_{1mix}^{'}-f_{2mix}^{'})\over m_D^2-m_{\eta 
'}}\biggr]~,
\nonumber\\
L^{\rho^+}\!\!\!&=&\!\!\!{1\over q^2-m_{\rho}^2+i\Gamma_{\rho}m_{\rho}}~,
\nonumber\\
L^{K^{*+}}\!\!\!&=&\!\!\!{1\over 2 g_{K^+}}~\biggl({g_{\rho}\over 
q^2-m_{\rho}^2+i\Gamma_{\rho}m_{\rho}}+{g_{\omega}\over 
3(q^2-m_{\omega}^2+i\Gamma_{\omega}m_{\omega})}+{2g_{\phi}\over 
3(q^2-m_{\phi}^2+i\Gamma_{\phi}m_{\phi})}\biggr)~,
\nonumber\\  
M^{D^0}\!\!\!&=&\!\!\!{g_{\rho}\over 
q^2-m_{\rho}^2+i\Gamma_{\rho}m_{\rho}}+{g_{\omega}\over 
3(q^2-m_{\omega}^2+i\Gamma_{\omega}m_{\omega})}~,
\nonumber\\
M^{D^+}\!\!\!&=&\!\!\!-{g_{\rho}\over 
q^2-m_{\rho}^2+i\Gamma_{\rho}m_{\rho}}+{g_{\omega}\over 
3(q^2-m_{\omega}^2+i\Gamma_{\omega}m_{\omega})}~,
\nonumber\\
M^{D_s^+}\!\!\!&=&\!\!\!-{2g_{\phi}\over 
3(q^2-m_{\phi}^2+i\Gamma_{\phi}m_{\phi})}~,
\nonumber\\
{\rm and} \nonumber\\
N\!\!\!&=&\!\!\!{g_{\rho}^2\over 
q^2-m_{\rho}^2+i\Gamma_{\rho}m_{\rho}}-{g_{\omega}^2\over 
3(q^2-m_{\omega}^2+i\Gamma_{\omega}m_{\omega})}-{2g_{\phi}^2\over 
3(q^2-m_{\phi}^2+i\Gamma_{\phi}m_{\phi})}~.
\end{eqnarray}
The  functions $f_{1mix}$, $f^{'}_{1mix}$, $f_{2mix}$ and $f^{'}_{2mix}$ 
are defined by 
\begin{eqnarray}
\label{mix}
f_{1mix}&=&{f_{\eta}\over\ \sqrt{8}}[{1+c^2\over f_{\eta}}+{sc \over f_{\eta 
'}}], \quad
f'_{1mix}={f_{\eta '}\over \sqrt{8}}[{sc\over f_{\eta}}+{1+s^2 \over f_{\eta 
'}}], \\ 
f_{2mix}&=&{f_{\eta}\over \sqrt{8}}[{1-5c^2\over f_{\eta}}-{5sc \over f_{\eta 
'}}]~~~{\rm and} ~~~ 
f'_{2mix}={f_{\eta '}\over \sqrt{8}}[{-5sc\over f_{\eta}}+{1-5s^2 \over 
f_{\eta '}}],\nonumber 
 \end{eqnarray}
where $s=\sin \theta_P$, $c=\cos \theta_P$ and $\theta_P\sim 20^0$ is the 
$\eta-\eta '$ mixing angle. The values of the masses, decay constants and decay 
widths used are given in Table 2.\\

\begin{table}[ht]
\begin{center}
\begin{tabular}{|c|c|c||c|c|c||c|c|c|c|}
\hline
$H$ & $m_H$ & $f_H$ & $P$ & $m_P$ & $f_P$ & $V$ & $m_V$ & $g_V$ & $\Gamma_V$ \\
\hline
\hline
$D$ & $1.87$ & $0.21 \pm 0.04$ & $\pi$ & $0.14$ & $0.13$ & $\rho$ & $0.77$ & 
$0.17$ & $0.15$ \\
$D_s$ & $1.97$ & $0.24 \pm  0.04$ & $K$ & $0.50$ & $/$ & $K^*$ & $0.89$ & $0.19$ 
& $/$ \\
$D^*$ & $2.01$ & $0.21 \pm  0.04$ & $\eta$ & $0.55$ & $0.13 \pm 0.008$ & 
$\omega$ & $0.78$ & $0.15$&$0.0084$  \\
$D_s^*$ & $2.11$ & $0.24 \pm  0.04$ & $\eta'$ & $0.96$ & $0.11 \pm 0.007$ & 
$\phi$ & $1.02$ & $0.24$&$0.0044$\\
\hline
\end{tabular}
\end{center}
\caption{The pole masses $m$, decay constants $f$ and decay widths $\Gamma$ in 
$GeV$; the constants $g_V$ in $GeV^2$.}
\end{table}

{\bf ACKNOWLEDGMENTS}\\

The research of S.F. and S.P. was supported in part by the Ministry of Science 
of 
the Republic of Slovenia. 
The research of P.S. was supported in part by the Fund for 
Promotion of Research at the Technion. One of us (S.F.) thanks the Physics 
Department at the Technion for the warm hospitality during her stay there, 
where part of this work was done. 
S. F. and S. P. are very grateful to G. Kernel and B. Ker\v sevan for many  
useful discussions. 

\newpage

{\bf Figure Captions}\\

{\bf Fig. 1.} Skeleton diagrams of various {\it long distance contributions} to 
the 
decay $D \to V l^+l^-$ resulting from Eq. (\ref{factor}). The spectator diagrams 
of type $A_{Spec,\gamma}$ (see 
Eq. (\ref{factor})) are shown in Fig. 1a, the spectator diagrams of type 
$A_{Spec,V}$ are shown in Fig. 1b and the weak annihilation diagrams 
$A_{Annih}$ are shown in Fig. 1c.  Different diagrams are denoted by the roman 
numbers $I-VIII$. The square in each diagram denotes the weak transition due to 
the long distance Lagrangian ${\cal L}_{LD}$ (Eq. \ref{weak}). This Lagrangian contains a product of two weak currents, each denoted by a dot. \\

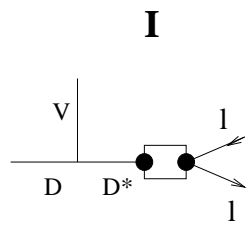
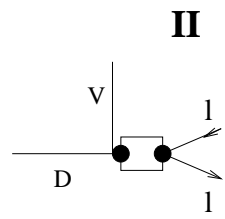
{\bf Fig. 2.} Skeleton diagrams of {\it short distance contributions} to the 
decay $D \to V l^+l^-$ due to $c\to ul^+l^-$ transition. The squares in the 
diagrams denote the weak transition due to the short distance Lagrangian ${\cal 
L}_{SD}$ (Eq. \ref{SD}). This Lagrangian contains a product of a quark and a 
lepton weak currents, each denoted by a dot. \\ 

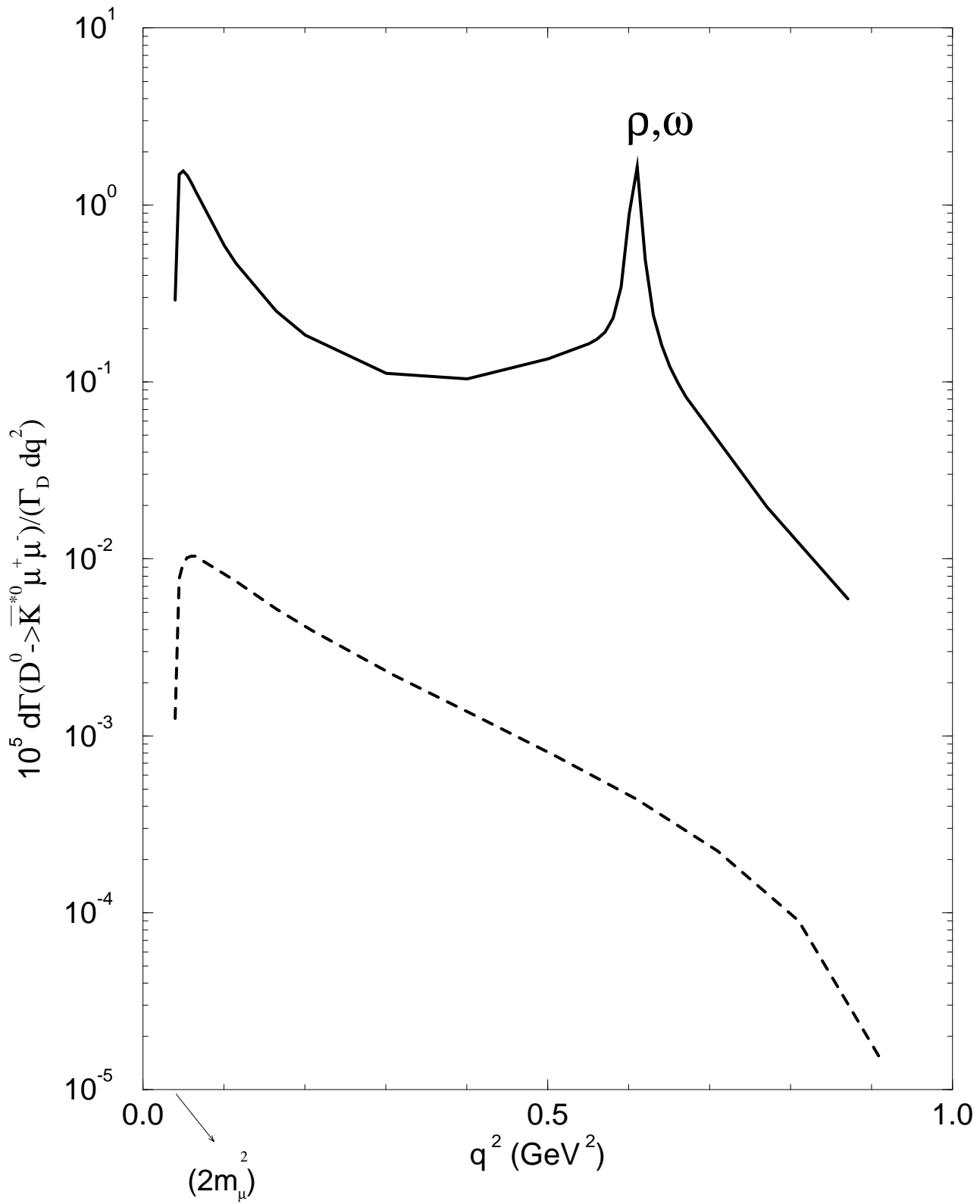
 {\bf Fig. 3.} The differential branching ratio 
$(1 /\Gamma_D ) d \Gamma (D^0 \to \bar K^{*0} \mu^+ \mu^-)/ d q^2$  
as a function of $q^2$ ($q^2$ the invariant $\mu^+ \mu^-$ mass). 
The full line corresponds to the total  branching ratio, while the  
dashed line  represents the nonresonant long distance part. In the calculation 
the model parameters $\lambda '= 0.26$ and $C_{VV\Pi}=0.31$ were used.\\

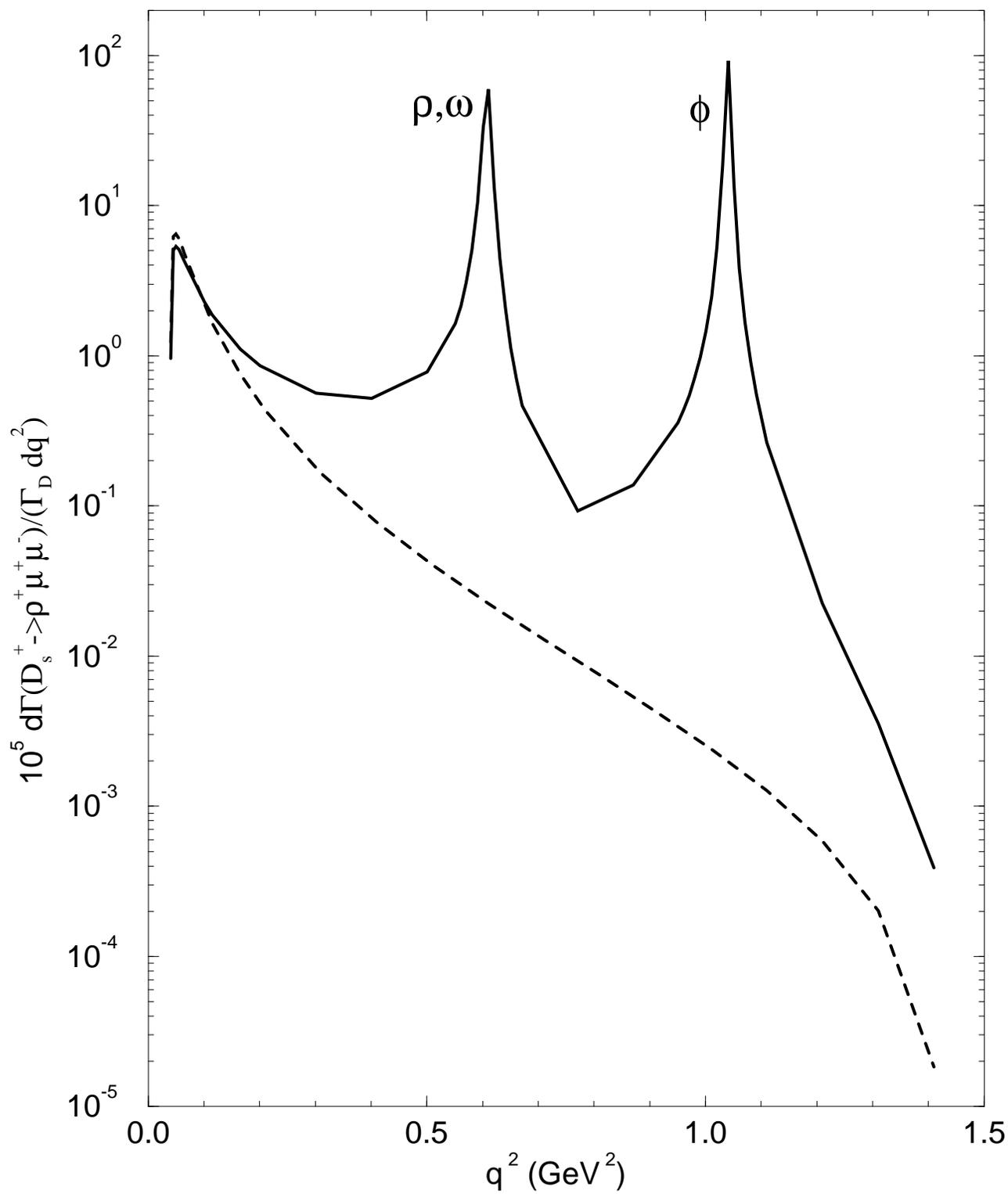
 {\bf Fig. 4.} The differential branching ratio 
$(1 /\Gamma_D ) d \Gamma (D^+_s \to \rho^{+} \mu^+ \mu^-)/ d q^2$  
as a function of $q^2$ ($q^2$ the invariant $\mu^+ \mu^-$ mass). 
The full line corresponds to the total  branching ratio, while the  
dashed line  represents the nonresonant long distance part. In the calculation 
the model parameters $\lambda '= 0.26$ and $C_{VV\Pi}=0.31$ were used.\\
 
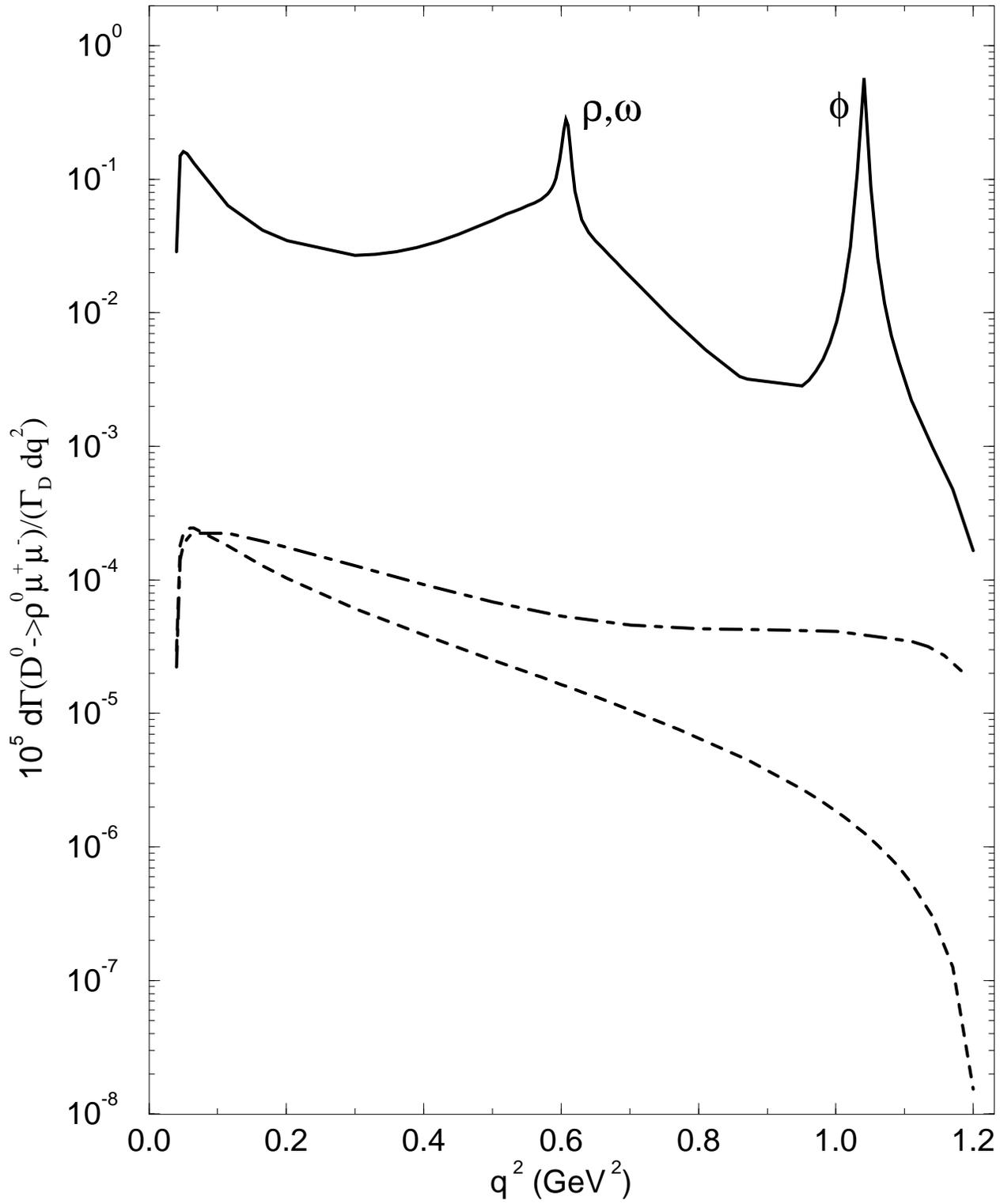
{\bf Fig. 5} The differential branching ratio 
$(1 /\Gamma_D ) d \Gamma (D^0 \to \rho^{0} \mu^+ \mu^-)/ d q^2$  
as a function of $q^2$ ($q^2$ the invariant $\mu^+ \mu^-$ mass). 
The full line represents to the total  branching ratio,  the  
dot-dashed line  represents the  short distance part, while the dashed line 
represents the nonresonant long distance part.  In the calculation the model 
parameters $\lambda '= 0.26$ and $C_{VV\Pi}=0.31$ were used.\\

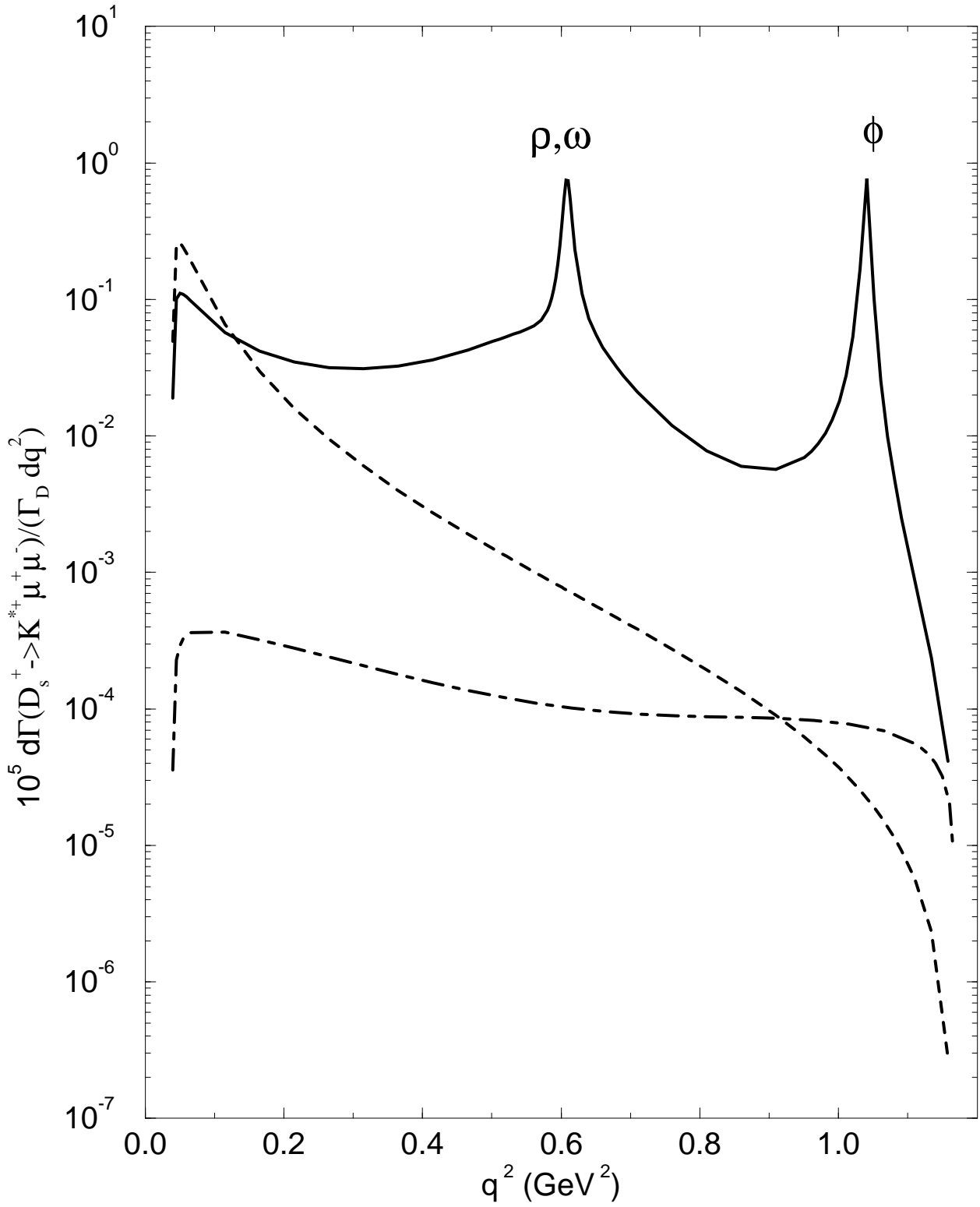
{\bf Fig. 6} The differential branching ratio 
$(1 /\Gamma_D ) d \Gamma (D_s^+ \to K^{*+} \mu^+ \mu^-)/ d q^2$  
as a function of $q^2$ ($q^2$ the invariant $\mu^+ \mu^-$ mass). The full line 
represents to the total  branching ratio,  the  
dot-dashed line  represents the  short distance part, while the dashed line 
represents the nonresonant long distance part.  In the calculation the model 
parameters $\lambda '= 0.26$ and $C_{VV\Pi}=0.31$ were used.\\

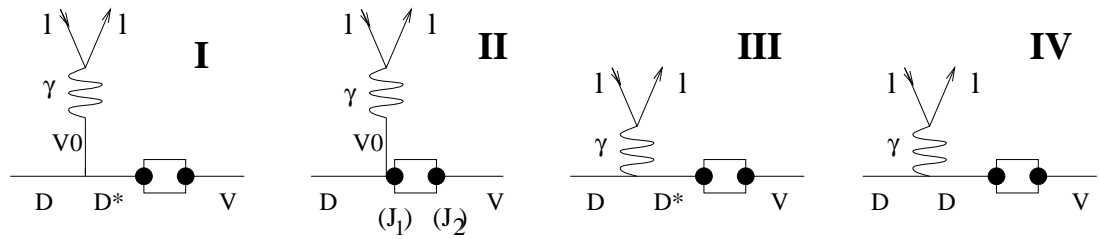
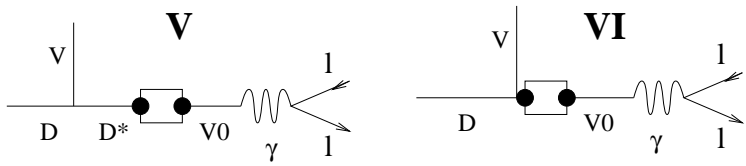
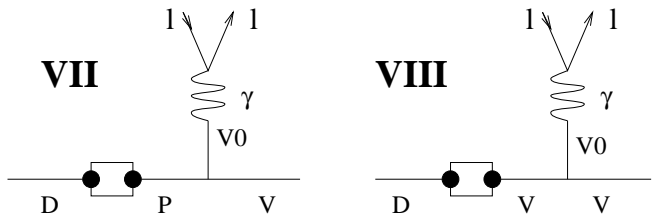
\end{document}